\documentclass[12pt,preprint]{aastex}
\shorttitle{The INTEGRAL/IBIS source AXJ1838.0-0655}
\shortauthors{Malizia et al.}

\begin{document}


\newcommand {\flux} {{$\times$ 10$^{-11}$ erg cm$^{-2}$ s$^{-1}$}}
\newcommand {\axj} {AX J1838.0-0655~}

\title{The INTEGRAL/IBIS source AXJ1838.0-0655: a soft X-ray to TeV 
$\gamma$-ray broad band emitter.\altaffilmark{1}}

\author{A. Malizia\altaffilmark{2}, L. Bassani\altaffilmark{2}, J. B. Stephen\altaffilmark{2}
A. Bazzano\altaffilmark{3}, P.Ubertini\altaffilmark{3}, A.J.Bird\altaffilmark{4}, 
A.J. Dean\altaffilmark{4}, V. Sguera \altaffilmark{4},M. Renaud\altaffilmark{5}, R. Walter\altaffilmark{6},
F. Gianotti\altaffilmark{2}}

\altaffiltext{1}{Based on observations obtained with the ESA science mission {\it INTEGRAL}}.
\altaffiltext{2}{IASF/INAF, Via Gobetti 101, I-40129 Bologna, Italy}
\altaffiltext{3}{IASF/INAF, Via Fosso del Cavaliere 100, I-00133 Rome, Italy} 
\altaffiltext{3}{School of Physics and Astronomy, University of
Southampton, Highfield, Southampton, SO 17 1BJ, UK.}
\altaffiltext{5}{Sap-CEA, Saclay, F-91191 Gif-sur-Yvette, France}
\altaffiltext{6}{{\it INTEGRAL} Science Data Centre,Chemin d'Ecogia 16,1291, Versoix, Switzerland}

\begin{abstract}
  We report on INTEGRAL observations of AX J1838.0-0655, one of
  the unidentified objects listed in the first IBIS/ISGRI survey
  catalogue and located in the Scutum arm region. This object,
  detected in the 20-300 keV band at a confidence level of
  15.3$\sigma$ (9 $\times$ 10$^{-11}$ erg cm$^{-2}$ s$^{-1}$)
  is the likely counterpart of the
  still unidentified TeV source HESS J1837-069. It has been
  detected in the past by various X-ray telescopes, including ASCA,
  implying that it is a persistent rather than a transient source; the
  ASCA image is compatible with the source not being resolved.  The
  broad 1-300 keV spectrum is characterized by an absorbed
  (N$_{H}$=6.7$\pm$1.3 $\times$ 10$^{22}$ cm$^{-2}$) and hard
  ($\Gamma$=1.5$\pm$ 0.2) power law continuum. Possible counterparts (radio and
  infrared) present within the X-ray error box are
  discussed, even if no clear association can be identified. The broad
  band spectrum together with the TeV detection suggests that \axj
  maybe a supernova remnant or a pulsar wind nebula, which has so far
  eluded detection in the radio band. This is  the second unidentified
  HESS source   that shows a substantial soft gamma-ray emission. 
\end{abstract}


\section{Introduction}
Within the sample of unidentified $\gamma$-ray sources found in the first
survey of our Galaxy, AX J1838.0-065 is particularly interesting due to its likely
physical association with the newly discovered TeV source HESS
J1837-069 (Aharonian et al. 2005a). In fact, its position coincides with the center of
gravity of the TeV emission and therefore has been suggested by
Aharonian et al. (2005a) as the most promising counterpart candidate
for HESS J1837-069.  Here, we report on INTEGRAL observations of \axj
showing substantial emission in soft $\gamma$-rays,
compare them  with previous X-ray measurements as well as with the TeV
data, describe the search for counterparts at other wavebands and
discuss the  possible nature of the source on the basis of our findings.

\section{The INTEGRAL data}
\axj is located in the Scutum arm region 
and has been independently detected as an INTEGRAL
source both in guest observer data by Molkov et al. (2004) and 
by Bird et al. (2004). The source has been detected up to 300 keV by ISGRI (Lebrun et al.2003), 
the low energy imager of IBIS (Ubertini et al. 2003). 
The observational data reported here are comprehensive of all observations 
 from revolution 46 (February 2003) to revolution 186 (April
2004) for a total exposure of 840 ksec. ISGRI images for
each available pointing were generated in 10 narrow energy bands using
the ISDC offline scientific analysis software OSA 4.1 (Goldwurm et al.
2003). Individual images were then combined to produce
a mosaic of a broader band to enhance the
detection significance using the procedure described in detail by Bird
et al. (2004). Figure 1 shows the 20-300 keV band image 
of the region surrounding the source which is detected with a
significance of 15.3$\sigma$ at a position corresponding to 
R.A.(2000) = 18h 38m 01.7s and Dec = -06$^{\circ}$ 54' 14.4'' with a positional uncertainty of $\leq$3'  
(Gross et al. 2003).
Fluxes for spectral analysis were extracted from fine band mosaics of
all revolutions added together and analysis was performed using  XSPEC v.11.2. A simple
power law provides a good fit to the IBIS data ($\chi^{2}_{\nu}$=0.6  for
5 d.o.f.) with  a photon index $\Gamma$=1.66$\pm$0.23 and  a
20-300 keV flux of 9 $\times$ 10$^{-11}$ erg cm$^{-2}$ s$^{-1}$.
The errors quoted  here and subsequently, correspond to 90${\%}$
confidence level for single parameter variation. \\
The flux of \axj for each individual pointing was also used to
generate the source light curve: no flares/variations are
visible in the light curve nor does the source show
periodicities/pulsations. Although the significance is low,
\axj seems to be a persistent source in the IBIS data. Dividing the 
observing period into 2 segments also indicates no variation in flux:
the detection significance remains at a level of 9 and 8$\sigma$ in
Mar-May 2003 and Sept-Oct 2003 respectively.\\

\section{The TeV high energy emission}
HESS J1837-069 is one of 10 sources found in the HESS survey of the
inner regions of the galactic plane and it is located at R.A.(2000)= 18h
37m 42.7s and Dec(2000)=-06$^{\circ}$ 55' 39'' with a positional
uncertainty in the range of 1-2' (Aharonian et al. 2005a).  
The quoted angular size is 4' although the TeV image (see figure 3 of Aharonian et al. 2005a)
suggests more extension and a non-Gaussian shape.
The statistical significance of the TeV detection is around 7-8$\sigma$ 
(depending if the source is taken as extended or point-like) with a source
flux above 0.2 TeV of 9 $\times$ 10$^{-12}$ photons cm$^{-2}$
s$^{-1}$: assuming a power law photon  index in the range 2-2.5 this value 
corresponds to 7-11 $\times$ 10$^{-12}$ erg cm$^{-2}$ s$^{-1}$ in the 0.2-10 TeV band.
HESS J1837-069 is at an angular distance of about 12' from the ASCA source G25.5+0.0
(Bamba et al. 2003), whose nature is still unknown although the X-ray
data indicate that it may be either a shell type  supernova remnant (SNR) or, 
alternatively, a pulsar wind nebula (PWN). Its position is, however, incompatible with
the TeV location. The nearest pulsar  to HESS J1837-069 is PSR J1837-06
at a distance of 7.4'.
This pulsar has a period of 1.9s, a distance of $\sim$5 kpc, a spin down age $\tau$ of 
39100 kyr and a spin down flux of 
1.5 $\times$ 10$^{-15}$ erg cm$^{-2}$ s$^{-1}$. Comparison between the 
HESS and the spin down luminosities indicates that this pulsar is not sufficiently energetic to power
the TeV radiation; furthermore its location is at the outskirts
of the emitting region. 
Aharonian and co-workers  (2005a) suggest that the  most likely candidate for HESS J1837-069 is
\axj  as it coincides with the centre of gravity of
the TeV source, even though  it is at the border of the extension 
which they quoted.
In figure 1, the green circle marks the
position and extension of HESS J1837-069 and indicates the clear spatial coincidence
between the IBIS/ISGRI object and the TeV source. \\
Although an EGRET object, 3EG J1837-0606/GEV1837-0610 (Hartman et al. 1999, Lamb \& Macomb 1997), 
is located nearby  ($\sim$ 50' away), its positional uncertainty ($\sim$11') is
such that an association with HESS J1837-069 is unlikely. 
Based on the 3rd EGRET catalogue (Hartman et al. 1999), the estimated upper limit above 100 MeV is  
$\sim$ 10$^{-7}$ photons cm$^{-2}$ s$^{-1}$.\\

\section{The X-ray source}

\axj was first detected in X-rays by the IPC instrument on the
Einstein satellite (2E4109/1E1835.3-0658,
Hertz and Grindlay 1988) at position R.A. (2000) = 18h 38m 04.2s and
Dec(2000) = -06$^{\circ}$ 55' 25.1'' with an associated 90$\%$ error radius
 of 49"
(cross in figure 1); the source flux was 0.014 counts in the
0.2-3.5 energy band
(or $\sim$ 3 $\times$ 10$^{-13}$  erg cm$^{-2}$ s$^{-1}$).\\
More recently, the source has been observed by ASCA in the 0.7-10 keV
range during the galactic plane survey (Sugizaki et al.  2001 and
Bamba et al. 2003).  Analysis of the ASCA-GIS images available in the
HEASARC database indicates that the source is point-like (3.6'
$\times$ 2.8' FWHM, against the instrument point spread function of 3')
although marginal extension on one side cannot be ruled out.  \axj
lies close ($\sim$20') to G25.5+0.0 and it is the brightest of 5 X-ray
sources found in the region (Bamba et al. 2003). The other 4 objects
are much weaker with a 0.7-10 keV flux below 4.1$\times$ 10$^{-12}$
erg cm$^{-2}$ s$^{-1}$ and none is detected by INTEGRAL;
one, AX J1837.3-0652, could be associated with PSR J1837-06
while AX J1838.1-0648 is likely the SNR 025.4-00.2.
The ASCA best fit position is at R.A.(2000) = 18h 38m 00.96s and Dec(2000) =
-06$^{\circ}$ 55' 55.20'' with an associated uncertainty of 1' in radius (white
circle in figure 1). From the positional coincidence it is evident
that the Einstein, ASCA and IBIS sources (see figure 1) are indeed the
same object, which is also likely to emit at TeV energies. 
We have analysed the ASCA-GIS spectrum obtained by combining two observations
performed in October 1997 and September 1999; the spectrum was made using
photons from a circular region of 3' radius and  background corrected
using data from a source free region of similar size. 
\axj is fairly bright in the 0.7-10 keV band with an integrated flux of
1.1 $\times$ 10$^{-11}$ erg cm$^{-2}$ s$^{-1}$ and a hard
($\Gamma$=0.8) and absorbed (N$_{H}$=4$\times$ 10$^{22}$ cm$^{-2}$ )
power law shape.  
The measured column density is in excess of the galactic value which in the
source direction  is 1.86 $\times$ 10$^{22}$ cm$^{-2}$ (Dickey $\&$ Lockman 
1990). \axj is
reported as being variable by Sugizaki et al. (2001) but Bamba et al. (2003)
do not confirm this variability.  
Our own inspection  of the ASCA data agrees with Bamba et al. (2003) findings that there is no evidence for variability. \\
The wide band spectrum, from 1 to 300 keV, using
the combined ASCA/INTEGRAL data, can be well fitted ($\chi^{2}_{\nu}$=1.3
for 19 d.o.f.) with an absorbed power law having a photon index of
$\Gamma$=1.5$\pm$0.2 and a total (galactic plus intrinsic) column
density of N$_{H}$=6.7$\pm$1.3 $\times$ 10$^{22}$ cm$^{-2}$.  To
account for a cross calibration mismatch between the two instruments
and/or source variability between the two observing periods, we have
introduced a constant in the fit, which when left free to vary 
provides a value of 1.1$^{+0.6}_{-0.4}$, indicating a good cross calibration
between the two instruments and, more importantly,
constancy in the flux between the two observing periods.  
The spectral energy distribution from X-rays to TeV $\gamma$-rays of \axj
is shown in figure 2 together with a detail of the
combined GIS-ISGRI spectrum.\\
We have also searched the HEASARC archive for the presence of the
source in the data of past X-ray missions and discovered an EXOSAT/ME
Galactic Plane survey source (GPS1835-070) located at a distance of 7' 
(Warwick et al. 1988): given the positional uncertainty of the
ME instrument (typically 9' for a 5$\sigma$ source) it is highly
likely that \axj was previously detected also by EXOSAT.  A
straightforward conversion from count rate to flux provides a
2-6 keV flux of 6.3 $\pm$ 0.9 $\times$ 10$^{-11}$ erg cm$^{-2}$
s$^{-1}$ to be compared with an ASCA flux in the same band of $\sim$
0.4 $\times$ 10$^{-11}$ erg cm$^{-2}$ s$^{-1}$; although this may be an
indication of flux variability, the lack of imaging capability of EXOSAT
suggests possible contamination from  nearby sources.  \axj is
not reported in the ROSAT all sky survey neither as a bright nor as a
faint object, however
with an upper limit which is compatible with the Einstein detection.\\
Browsing the BeppoSAX archive, we also found two PDS pointings (on
March and April 2001) devoted to the study of 3EG J1837-0606, which
contained serendipitously also \axj (Malizia et al. 2004). There is a
strong detection ($\sim$13$\sigma$) in the combined PDS data. 
Superposition of the PDS error box of 1.3 degrees on the
IBIS map indicates that the emission is likely dominated by \axj, but
a contribution from the nearby IGR J18410-0535 cannot be ruled out. If
fitted with a simple power law ($\chi^{2}$=7.8 for 8 d.o.f.) plus a
constant to account for the loss in sensitivity due to the distance of
\axj from the center of the PDS pointing ($\sim$44'), these data give a
photon index of $\Gamma$=2.5$^{+0.3}_{-0.3}$ and a 20-100 keV flux of
3.6 $\times$ 10$^{-11}$ erg cm$^{-2}$ s$^{-1}$, a similar flux but a
steeper spectral index than ISGRI. Although this is potentially
an interesting result, the lack of imaging capability, which is
crucial in the crowded regions of the galactic plane, suggests some
caution in interpreting the PDS (as well as EXOSAT) data. Instead, one
important conclusion is that \axj has always been detected above a few
keV whenever observed, strongly suggesting that it is a persistent X-ray
source.

\section{Search for counterparts at other wavebands}
The visual extinction in the \axj direction is
A$_v$=36.3 (from the relation by Predehl and Schmitt (1995)).  
This implies that any optical
counterpart will be extremely weak or invisible and that the search
for identification should be carried
out at radio and/or infrared frequencies.
Inspection of various databases provides indications for the presence
of a strong radio source (14 $\sigma$) at the border of the ISGRI
error box, but outside the smaller X-ray positional uncertainty of 
ASCA/Einstein: TXS1835-069-CUL1835-06, associated in Simbad with a
candidate supernova remnant, SNR025.3-00.1 (Dulk and Slee 1972).  This source has a
size of $\sim$ 5' and the radio spectrum is a power law
(S($\nu$)=A$\nu^{\alpha}$) of $\alpha$ (energy index) equal to -1 in
the 0.08-5 GHz range.  At 1.4 GHz the source morphology is complex as
TXS1835-069 is resolved into a point-like object and a linear double
(Helfand et al. 1998) resulting in 3 objects listed in the NVSS 
(NRAO VLA Sky Survey) catalogue and displayed in figure 3a. 
As shown in the figure  all three objects are 
inside the ISGRI positional uncertainty but outside the X-ray 
error boxes. The extended region north-west of \axj coincides with 
a star forming complex containing yet another supernova remnant 
SNR025.4-00.2 and few  giant HII regions.
Helfand and co-authors estimate for the double source a spectral index of
$\alpha$$\ge$ -0.4, while the point source has a slightly inverted
spectrum at low frequencies and then declines above 1.4 GHz.  They
also suggest that TXS1835-069 is likely composed of a pair of
foreground extragalactic objects and is not a supernova.  Clearly deeper
radio observations of this radio complex are required to definitely
identify its true nature and eventually rule out (or in) any connection
with \axj/HESS J1837-069.
Inspection of the NVSS image indicates a flux limit of 1-2 mJy at
1.4 GHz in this region (Condon et al. 1998), implying for \axj a very
low radio brightness compared to other TeV sources (Helfand et al. 2005) if no connection with TXS1835-069 will be found.\\
The X-ray positional uncertainty  
is too large to allow a proper search for counterparts at
infrared wavebands; this is also true for the sky region which is in
common to the 3 error boxes (figure 3b), i.e. the most likely to
contain the counterpart of \axj. Of the many infrared candidates, one
is particularly interesting as it is detected at all infrared
wavelengths from 1.25 to 21.34 $\mu$m: G025.2454-00.1885 which is the
only MSX6C (Midcourse Space Experiment) object detected in the X-ray
error box at R.A.(2000) = 18h 38m 01.61s and 
Dec(2000) = -06$^{\circ}$ 55' 23.5''
(see figure 3b which is a 8.3 $\mu$m image of the region containing AX J1838.0-0655). 
This source is fairly bright in the infrared (1.8 Jy at 8.3 $\mu$m and
$\sim$0.02 Jy in J band), but not in the optical where the B magnitude
is likely to be fainter than 19-20. 
It is unclear at this stage what is the nature of this infrared source 
but its brightness argues against too much reddening thus questioning the  
association with \axj.

\section{What is the nature of \axj?}
\axj is clearly a persistent and possibly non variable high energy
source.  The X/$\gamma$-ray spectrum is hard and absorbed by a column
density in excess of the galactic value; 
the fact that we see no turn off in the soft $\gamma$-ray 
emission up to at least 100-300 keV is very puzzling 
although it must be noted that 
the sensitivity of ISGRI at these high energies requires 
longer exposures to get 
more statistically sound results.
The X-ray morphology indicates that
\axj is more likely to be a point source rather than extended, while  
the TeV results are more
compatible with an extended region of emission. 
Many of these features are indicative of non-thermal radiation and strongly
suggest a classical TeV emitter scenario, i.e. a  SNR and/or a
PWN (Aharonian et al. 2005a) or alternatively a binary system 
such as PSR B1259-63 (Aharonian et al. 2005b).
This last possibility does not fit with the lack  of strong
X/$\gamma$-ray variability and  
the extension of the source at TeV energies. 
On the other hand, the constant flux and  X-ray/TeV  morphology are more in favour 
of a shell type SNR and/or PWN scenario.
Note, however, that in a shell type SNR a cut off in the X-ray spectrum 
is expected to be well below 100 keV (Uchiyama et al. 2003) assuming a 
standard synchroton radiation model.
Detection of both types of SNR in the X-ray band but not at radio
frequencies is intriguing.  However, if the SNR magnetic field is
weak and/or the source is far away,  then the radio emission may be too faint to be detected in a shell type SNR
(Bamba et al. 2003 and references therein). Also for PWNs, the ratio between the
luminosity in X-ray and that in radio is still an open issue: there
are sources of this type with X-ray luminosities similar to the Crab but
with  radio fluxes 2 to 3 order of magnitude weaker (Slane et al.
2004).  Finally, Helfand et al. (2005) have recently noticed a large range of more than 3 orders of magnitude
in the radio to TeV flux ratios and a preference for TeV sources to show
X-ray synchrotron emission.
Thus it may not be surprising to find both SNR/PWN type of objects
first in X-rays and subsequently at radio frequencies where they may
be still too faint to have been picked up by current radio surveys.  The
possible association of \axj with the MSX source is also unusual.
This object is certainly related to thermal emission
and so difficult to recoincile with \axj characteristics.
Alternatevely it may indicate a totally different scenario to explain
the TeV emission: for example, the interaction of a molecular cloud
(likely to be detected by the MSX instrument) either with an expanding
SNR shell or with the stellar winds from an OB star association are
both viable sites of TeV emission. In this case, however, 
 it would be more difficult to
produce the X/$\gamma$-ray radiation. Also the vicinity of the star 
forming complex 
visible both in radio and far-infrared is interesting as it can provide 
the ambient starlight photons
for inverse Compton scattering of the synchrotron emitting high energy
electrons. Bamba et al. (2003), 
estimate a distance of $\sim$ 8 kpc for this region, a  value which 
is consistent with the fact that the source is located in the Scutum
arm (Koyama et al. 1991).  For this distance, we have 
X-ray, soft gamma and TeV  luminosities of 
10, 32 and 5.7$\times$ 10$^{34}$ erg
s$^{-1}$ respectively; these luminosities are similar to the values observed
in the few HESS sources which have been clearly identified with PWNs
or shell type SNRs (see Table 1).
It is interesting to note that the observed values are
closer to the first type of objects rather than to the second one.
Overall, we can conclude that all
observational evidences suggest at this stage that \axj may well be a
shell type SNR or a PWN.

\acknowledgments 
We would like to thank Dr A. Bamba for the ASCA data.
We acknowledge the following funding Agencies: in
Italy, Italian Space Agency financial and programmatic support via
contracts I/R/046/04.

\clearpage

\begin{figure}
\epsscale{.80}
\plotone{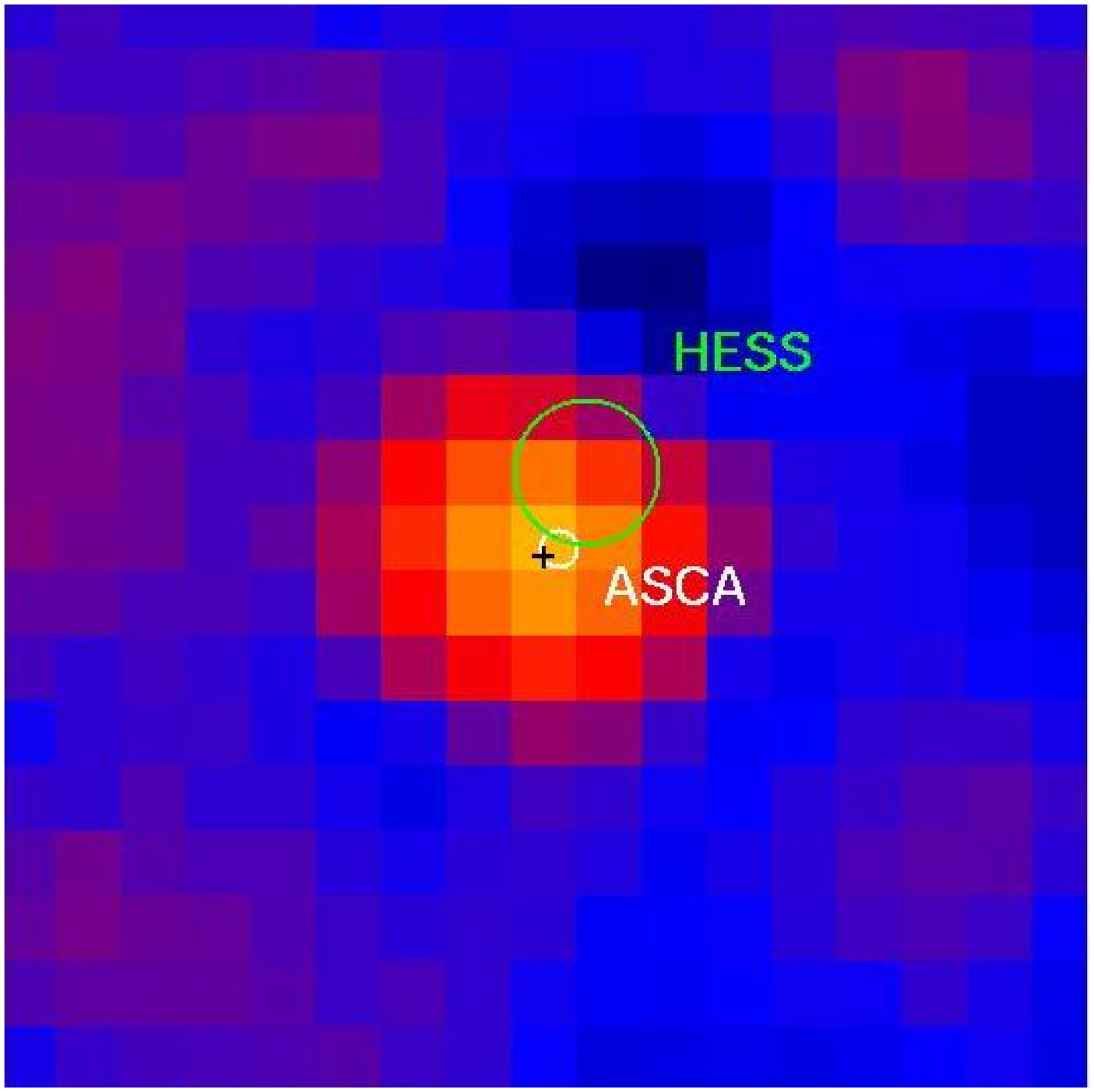} \caption {The IBIS/ISGRI 20-300 keV significance 
map (1 $\times$ 1 degree) showing the location of AX J1838.0-0655 as well as the position and 
extension of HESS J1837-069 (green circle), the ASCA position and uncertainty (white circle) and
the Einstein position (black cross). \label{fig1}}
\end{figure}

\clearpage

\begin{figure}
\plotone{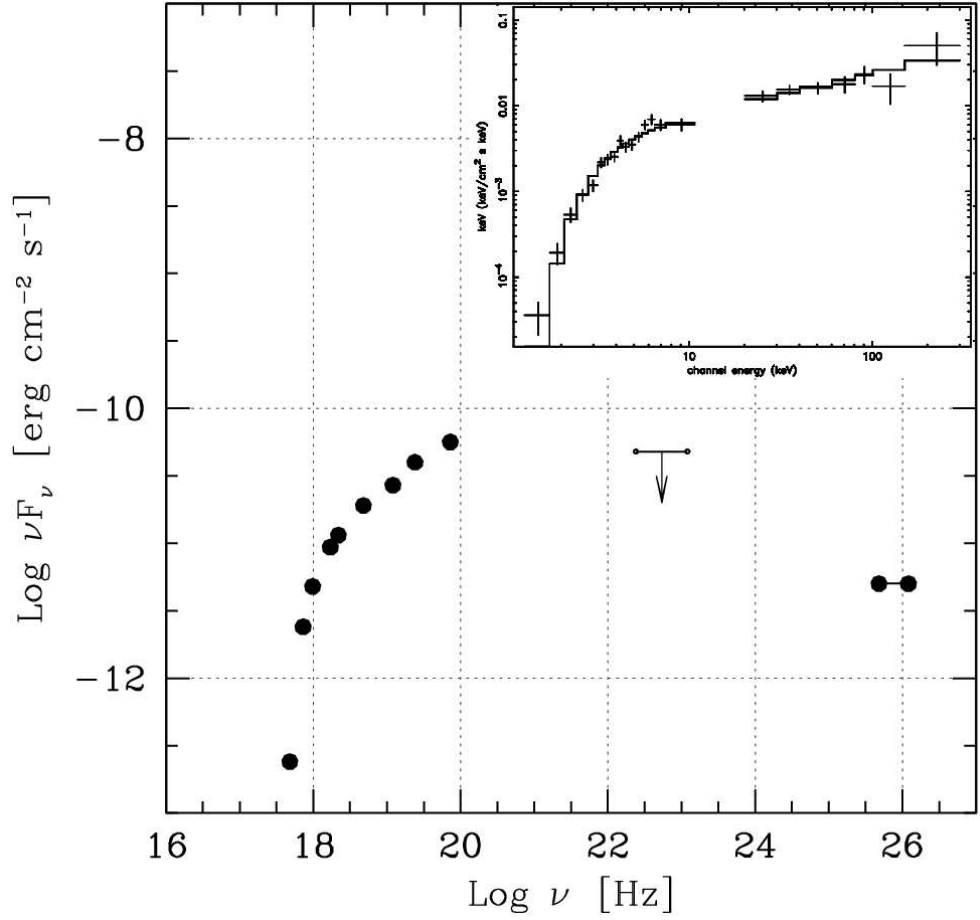} \caption{Spectral Energy Distribution of AX J1838.0-0655 from X-rays to TeV
gamma-rays including EGRET upper limit. The insert is the blow-up of the combined ASCA-GIS/INTEGRAL-ISGRI spectrum.\label{fig2}}
\end{figure}

\clearpage

\begin{figure}
\epsscale{.80}
\plotone{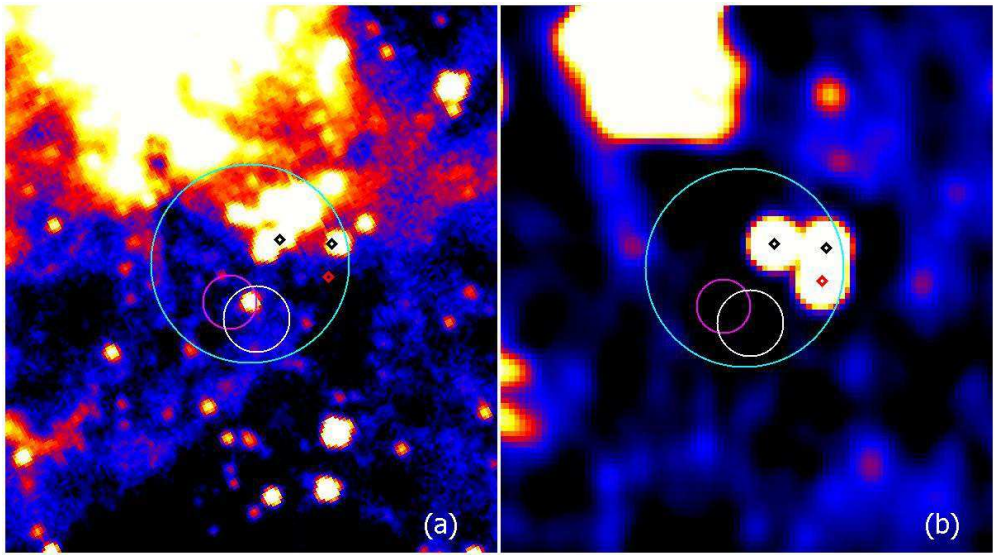} \caption{(a) NVSS image of the region surrounding AX J1838.0-0655.
The circles are from bigger to smaller: ISGRI, ASCA, Einstein. Diamonds are the positions of the 
radio source complex discussed in the text. The extended emission north of AX J1838.0-0655 is a 
star forming complex containing giant HII regions and the SNR 025.4-00.2. (b) MXS image at 8.3$\mu$m of the same
sky area. Symbols are the same as described in (a).\label{fig3}}
\end{figure}

\clearpage

\begin{table}
\begin{center}
\caption{X-ray, soft gamma-ray and TeV luminosities$^{a}$}
\begin{tabular}{crrrrrc}
\tableline\tableline
Source   &  2-10 keV   &  20-100 keV  & 0.2-10 TeV &  D(Kpc)& TeV type\\
\tableline
AX J1838.0-0655    & 10.0  (1) &  32.0 (1)  &  5.7(1)   & 8    &  ???\\
MSH15-52           & 8.0(2)    &  47.0 (3)  & 10.0(4)   & 5    &  PWN\\
G0.9+0.1           & 5.0(5)    & $<$61.0 (3)  &  4.0(6)   & 8.5  &  PWN\\
RX J1713.7-3946    & 3.3(7)    & $<$0.3  (3)  &  0.9(8)   & 1    &  SNR\\
RX J0852.0-4622    & 0.02(9)   & $<$0.01 (3)  &  0.05(10) & 0.2  &  SNR\\
\tableline
\end{tabular}
\tablenotetext{a}{in units of 10$^{34}$ erg s$^{-1}$ }

(1) this work; (2) Gaensler et al. 2002, (3) Bird et al. 2005; (4)
Aharonian et al. 2005c; (5)Porquet, Decourchelle $\&$ Warwick 2003; 
(6) Aharonian et al. 2005d;(7) Slane et al. 1999, (8) Uchiyama et al. 2005;
(9)Slane et al. 2001; 810) Aharonian et al. 2005e. 
\end{center}
\end{table}
 

\begin{thebibliography}{}
\bibitem{} Aharonian, F.; et al., 2005a, Science, 307,1938
\bibitem{} Aharonian, F.; et al., 2005b, submitted to A\&A
\bibitem{} Aharonian, F.; et al., 2005c, A\&A 435, L17
\bibitem{} Aharonian, F.; et al., 2005d, A\&A 432, L25
\bibitem{} Aharonian, F.; et al., 2005e, A\&A in press (astroph.0505380)
\bibitem{}
Bamba, A; Ueno, M.; and Koyama, K. 2003, ApJ, 589, 253
\bibitem{}
Bassani, L.; Malizia, A.; Stephen, J. B. et al. 2004, Atel \#232
\bibitem{}
Bird, A. J.; Barlow A.J.; Bassani L.; et al. 2004, Ap.J.,607, L33 
\bibitem{}
Bird, A. et al. 2005 in preparation
\bibitem{}
Condon, J. J.; Cotton, W. D.; Greisen, E. ;  et al. 1998, AJ, 115, 1693
\bibitem{}
Dulk, P.$\&$  Slee, O. 1972, AuJPh, 25, 429
\bibitem{}
Gaensler B.M.; Arons,J.; Kaspi,V.M;   et al. 2002,Ap.J. 569,878
\bibitem{}
Goldwurm, A., David, P., Foschini, L. et al. 2003, A\&A, 411, L223
\bibitem{}
Gros, A.; Goldwurm, A.; Cadolle-Bel; M.; et al. 2003, A\&A 411, 179
\bibitem{}
Hartman, R. C.; Bertsch, D. L.; Bloom, S. D; et al. 1999, Ap.J.S., 123 79
\bibitem{}
Helfand, D.J.; Becker, R. H.; White, R. L. 2005, Ap.J. submitted
\bibitem{}
Helfand, D.J.; Velusamy, T.; Becker, R.H.; Lockman, F.J. 1998, ApJ, 341,151
\bibitem{}
Hertz, P. \& Grindlay, J. 1988, AJ, 96
\bibitem{}
Koyama, K.; Kunieda, H.; Takeuchi; Y., Tawara, Y. 1991, Ap.J, 370, L77
\bibitem{}
Lamb, R. C.$\&$ Macomb, D. J. 1997, Ap.J., 488, 872
\bibitem{}
Lebrun, F.; Leray, J.P.; lavocat, P.; et al., 2003, A\&A, 411, L141
\bibitem{}
Malizia, A.; Bassani, L.; Landi; R.; et al. 2004, 
Proceedings of the V INTEGRAL 
Workshop, Munich 16-20 February 2004. ESA SP-552
\bibitem{}
Molkov, S. V.; Cherepashchuk, A. M.; Lutovinov, A. A.; et al. 2004,  
Astr. Lett., 30, 534
\bibitem{}
Predehl, P. $\&$ Schmitt, J.H.M.M. 1995, A\&A, 293, 889
\bibitem{}
Porquet,D.; Decourchelle, A.; $\&$ Warwick,R.S. 2003, A\&A, 401, 197
\bibitem{}
Slane ,P.;Gaensler,B.M.; Dame,T.M.;et al. 1999,Ap.J., 525, 357
\bibitem{}
Slane, P.;Hughes, J.P.; edgar, R.J.; et al. 2001,Ap.J., 548, 814
\bibitem{}
Sugizaki, M.; Mitsuda, K.; Kaneda, H.; et al. 2001, Ap.J.S., 134, 77 
\bibitem{}
Ubertini P.; Lebrun F.; Di Cocco G.; et al. 2003 A\&A 411, L131 
\bibitem {}
Uchiyama, Y.; Aharonian, F. A.; Takahashi, T. 2003  A\&A 400, 567
\bibitem{} 
Uchiyama, Y.; Aharonian, F.A.; Takahashi, T.; et al. 2005, Proc.of Int. Symp.
on High energy gamm-ray Astronomy, Heidelberg 2004 (astroph.0503199)
\bibitem{} 
Warwick, R.S.; Norton, A.J.; Turner, M.J.L.; Watson, M.G.; Willingale, R. 
1988, MNRAS, 232, 551
\end{thebibliography}
\end{document}